\documentclass[cameraready]{Interspeech}
\title{wav2tok 2.0: Scalable Audio Tokenization Maintaining Explicit Pairwise Token Alignment for Efficient Audio Retrieval}

\author[affiliation={1}, orcid=0009-0005-2535-2883]{Adhiraj}{Banerjee}
\author[affiliation={1,2}, orcid=0000-0002-1207-1258]{Vipul}{Arora}
\address{$^1$ Department of Electrical Engineering, Indian Institute of Technology Kanpur, India \\
        $^2$ Department of Electrical Engineering, KU Leuven, Belgium }

\email{adhirajbanerjee35@gmail.com, vipular@iitk.ac.in, vipul.arora@kuleuven.be}


\usepackage{comment}
\usepackage{multirow}
\begin{document}

\maketitle


\begin{abstract}
Learning discrete speech representations that preserve similarity across variable-length utterances is central to query-by-example spoken term detection (QbE-STD). While wav2tok introduced CTC-based sequence alignment to enforce token consistency, its tightly coupled clustering and alignment training recipe limits scalability. We propose wav2tok 2.0, a scalable alignment-aware speech tokenizer built on the BEST-STD backbone. wav2tok 2.0 employs staged training, first learning discriminative, speaker-invariant representations via contrastive learning and vector quantization, and then enforcing pairwise token consistency using a CTC alignment loss and a novel DTW-aligned framewise prediction objective with adaptive weighting. Experiments show that wav2tok 2.0 consistently outperforms BEST-STD and general-purpose tokenizers on QbE-STD while remaining efficient and scalable.
\end{abstract}

\section{Introduction}

Query-by-example spoken term detection (QbE-STD) aims to retrieve utterances from
large audio archives that contain a given spoken query, operating directly on
audio signals rather than text, and is central to applications such as audio
indexing, podcast retrieval, and voice search
\cite{alberti2009audio,ogata2009podcastle,wang2008introduction}. Early approaches
relied on ASR-based representations using phone or grapheme lattices
\cite{mamou2007vocabulary,miller2007rapid,wang2008comparison}, but depend on highly
accurate recognition, which is challenging for short queries, unseen words, and
mismatched acoustic conditions
\cite{saraclar2004lattice,can2011lattice}. DTW-based template matching avoids
explicit transcription by directly aligning acoustic sequences
\cite{ram2020neural,ram2018cnn,tsai2021segmental}, but is computationally
impractical at scale. Embedding-based methods improve efficiency by learning
acoustic word representations
\cite{chung2016unsupervised,he2016multi,kamper2016deep,chen2018phonetic,hu2021acoustic,banerjee2023enc},
yet typically require reliable word segmentation, while fingerprinting and
hashing approaches remove this requirement but remain sensitive to speaker and
channel variability
\cite{singh2022attention,singh2023simultaneously,singh2024flowhash}.

Recent advances in speech tokenization offer a complementary perspective.
General-purpose tokenizers such as HuBERT~\cite{hsu2021hubert},
WavLM~\cite{chen2022wavlm}, SpeechTokenizer~\cite{zhang2023speechtokenizer}, and
EnCodec~\cite{defossez2022high} learn discrete representations via self-supervised
predictive or reconstruction objectives, enabling compact representations for
recognition, synthesis, compression, and generation, but are not optimized for
retrieval-oriented alignment. Within STD, BEST-STD~\cite{singh2025best} shows that
scalable, speaker-invariant tokenization can be learned using contrastive
learning and vector quantization, where alignment is handled implicitly via
DTW-based positive sampling. Subsequent extensions improve robustness and
scalability—LAST-STD~\cite{singh2025language} via optimal transport–based codebook
balancing in multilingual settings and BEST-STD~2.0~\cite{singh2025best2} under
noisy conditions
\cite{cuturi2013sinkhorn}—but retain implicit alignment mechanisms.

In contrast, wav2tok~\cite{banerjee2022wav2tok} was the first to explicitly cast
speech tokenization as a sequence alignment problem within a pairwise learning
framework.  It explicitly enforces edit-distance preservation between token sequences using
a CTC-based \cite{graves2012connectionist} likelihood objective, providing a principled alignment mechanism
between paired utterances.
Despite its conceptual appeal, wav2tok \cite{banerjee2022wav2tok} relies on repeated clustering steps and a
tightly coupled contrastive–alignment training recipe, which makes it
difficult to scale beyond relatively small datasets.

In this paper, we introduce \emph{wav2tok 2.0}, a scalable reformulation that
combines the strengths of BEST-STD \cite{singh2025best} and wav2tok \cite{banerjee2022wav2tok}.
wav2tok 2.0 adopts the BEST-STD \cite{singh2025best} architecture to learn robust and discriminative
frame-level representations, while reintroducing explicit token-level alignment
constraints in a scalable manner.
Training is performed in two stages: (i) representation learning via
self-supervised contrastive objectives and vector quantization, and (ii) pairwise
token alignment using CTC-based alignment together with a novel
DTW-aligned framewise token prediction loss. stabilized via adaptive weighting. Experiments show that
wav2tok 2.0 consistently outperforms BEST-STD \cite{singh2025best} and contemporary tokenizers on
QbE-STD benchmarks without compromising scalability. Code: {\scriptsize \url{https://github.com/adhiraj69/wav2tok2}}.

\section{Method}
\label{sec:method}

\subsection{Encoder and Tokenization Backbone}

wav2tok 2.0 adopts the BEST-STD \cite{singh2025best} architecture as a backbone for scalable
representation learning.
The encoder $f_\theta$ maps an input utterance $x$ to a sequence of frame-level
embeddings $\mathbf{Z}=\{z_t\}_{t=1}^{T}$ using a spectrogram frontend followed
by a bidirectional Mamba-based \cite{gu2021efficiently} state-space model.
The final encoder outputs are projected to a $d$-dimensional space and
$\ell_2$-normalized.
Each frame embedding is discretized using a vector quantizer with codebook
$\mathcal{C}=\{c_k\}_{k=1}^{K}$.
Since both embeddings and centroids are $\ell_2$-normalized, token assignment is
equivalently defined via cosine similarity:
$q_t = \arg\max_{k} \; z_t^\top c_k $.
The codebook is trained using exponential moving average updates with centroid
normalization, as in BEST-STD \cite{singh2025best}.
\begin{figure*}
  \centering
  \includegraphics[width=\linewidth]{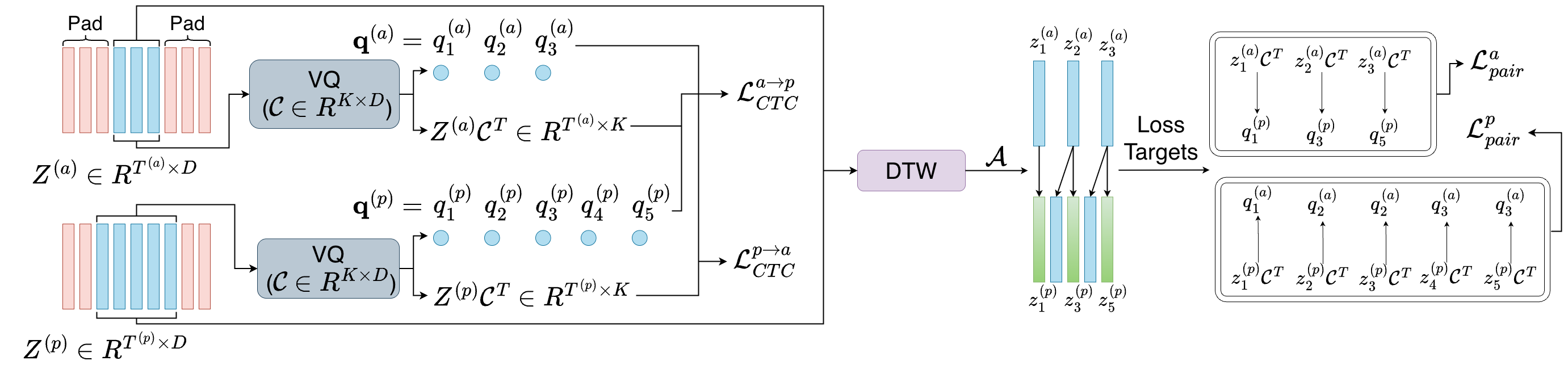}
  \caption{Stage II pairwise alignment framework combining CTC-based sequence alignment with a novel DTW-aligned framewise token prediction. }
  \label{fig:stageII_DTW}
%\vspace{-12pt}
\end{figure*}

\subsection{Stage I: Discriminative Pretraining}
In the first stage, the encoder and codebook are trained to produce a
discriminative latent space using the self-supervised learning framework of
BEST-STD \cite{singh2025best}.
Given paired utterances $(u^{(a)},u^{(p)})$ of the same term, which may have
different durations, the encoder produces embedding sequences
$\mathbf{Z}^{(a)}=\{z^{(a)}_t\}_{t=1}^{T^{(a)}}$ and
$\mathbf{Z}^{(p)}=\{\tilde{z}^{(p)}_t\}_{t=1}^{T^{(p)}}$.
DTW is used to align these variable-length sequences,
yielding a monotonic many-to-many correspondence between frames.
Frame-level anchor–positive pairs are then formed by selecting, for each anchor
frame $t$, the aligned positive frame with maximum cosine similarity.

The encoder is trained using a SimCLR-style \cite{chen2020simple} contrastive objective
$\mathcal{L}_{\text{contrast}}$ together with a commitment loss
$\mathcal{L}_{\text{commit}}$ that encourages consistency between embeddings and
their assigned centroids.
This stage is necessary to obtain a stable and well-separated latent
space; without it, the token alignment objectives introduced in Stage~II
become ill-conditioned.

\subsection{Stage II: Pairwise Token Consistency Learning}

wav2tok 2.0 introduces token-level consistency constraints between paired
utterances, building upon the pairwise alignment formulation of wav2tok \cite{banerjee2022wav2tok} and
extending it with a finer-grained alignment objective.
\subsubsection{CTC-based pairwise alignment.}
Following wav2tok \cite{banerjee2022wav2tok}, we enforce pairwise consistency by \emph{maximizing the
likelihood} of the token sequence obtained from one utterance under the
frame-level representations of the paired utterance.
Specifically, given anchor representations $\mathbf{Z}^{(a)}$ and the token
sequence $\mathbf{q}^{(p)}$ extracted from the positive utterance, we define
token posteriors using cosine similarity between $\ell_2$-normalized embeddings
and codebook centroids:
\begin{equation}
\log p(q = k \mid z_t)
=
\log \mathrm{Softmax}\big(z_t^\top c_k\big),
\qquad k \in \{1,\dots,K\}.
\end{equation}

Since vector quantization produces repeated tokens across adjacent frames, we
apply a deduplication operator $\mathrm{dedup}(\cdot)$ to obtain a compressed
token sequence with no consecutive repetitions.
We then maximize the likelihood of this deduplicated sequence under the anchor
representations using the CTC forward–backward algorithm:
\begin{equation}
\mathcal{L}_{\text{CTC}}^{a \rightarrow p}
=
\mathrm{CTC}\big(
\log p(\mathbf{q} \mid \mathbf{Z}^{(a)}),
\mathrm{dedup}(\mathbf{q}^{(p)})
\big).
\end{equation}

In contrast to standard CTC formulations \cite{graves2012connectionist}, we explicitly disallow blank
transitions by assigning the blank symbol a log-probability of $-\infty$.
This is appropriate in our setting because the deduplicated token sequences
contain no consecutive repetitions; the blank symbol—typically used to model
token repetition and separation—is therefore unnecessary. This also implicitly removes explicit repeat transitions at the token level.
Frame-level repetitions are instead handled implicitly through the many-to-one
alignment paths marginalised by the CTC forward–backward recursion.
The same objective is applied symmetrically for $p \rightarrow a$ and averaged.

\subsubsection{DTW-aligned framewise token prediction (proposed).}
In addition to sequence-level alignment, we introduce a framewise token
prediction loss that exploits the DTW alignment path.
Let $\mathcal{S}(t)$ denote the set of frames in $\tilde{\mathbf{Z}}$ aligned to
anchor frame $t$.
We define a frame-level token target as
\begin{equation}
\tilde{q}_t^{(a)} =
q^{(p)}_{\arg\max_{j \in \mathcal{S}(t)}
\; z_t^{(a)\top} z_j^{(p)}} .
\end{equation}

Using the same cosine-similarity-based token posteriors, we minimize a
negative log-likelihood loss
\begin{equation}
\mathcal{L}_{\text{pair}}^{a}
=
-\sum_{t} \log p(\tilde{q}_t^{(a)} \mid z_t^{(a)}),
\end{equation}
with a symmetric term $\mathcal{L}_{\text{pair}}^{p}$ defined analogously.
This objective enforces fine-grained cross-view agreement between frame-level
representations and discrete token assignments under monotonic DTW alignment.
Figure~\ref{fig:stageII_DTW} illustrates the Stage~II pairwise alignment framework, combining CTC-based sequence alignment with a novel DTW-aligned framewise token prediction loss.

\subsection{Training Objective}

The full Stage~II training objective is defined as
\begin{equation}
\mathcal{L}
=
\alpha_{\text{ctr}} \mathcal{L}_{\text{contrast}}
+
\alpha_{\text{com}} \mathcal{L}_{\text{commit}}
+
\lambda_{\text{CTC}} \mathcal{L}_{\text{CTC}}
+
\alpha_{\text{pair}} \mathcal{L}_{\text{pair}},
\end{equation}
where $\alpha_{\text{ctr}}$, $\alpha_{\text{com}}$, and $\alpha_{\text{pair}}$
are fixed cost factors (set to $1.0$ in all experiments) controlling the relative contributions of the contrastive,
commitment, and DTW-aligned framewise prediction losses, respectively.

The CTC-based alignment objective operates on a different scale than the
frame-level losses and is particularly sensitive to the quality of the latent
representation.
To prevent the CTC term from dominating optimization or becoming numerically
unstable, we employ an adaptive weighting strategy.
Specifically, at each training iteration we scale the CTC loss as
\begin{equation}
\lambda_{\text{CTC}}
=
\gamma \,
\frac{\mathcal{L}_{\text{contrast}}}{\mathcal{L}_{\text{CTC}} + \varepsilon},
\end{equation}
where $\gamma$ is a constant (set to $0.5$ in all experiments) and $\varepsilon$
is a small constant for numerical stability.
This normalization maintains the effective magnitude of the CTC term at
approximately half that of the contrastive loss throughout training, ensuring
stable convergence while preserving the alignment signal.

Finally, Stage~I pretraining plays a critical role in stabilizing this objective:
by establishing reliable clustering of the frames before 
token-level alignment, the CTC forward--backward recursion remains well-defined
and numerically stable.

\section{Experiments}
\label{sec:experiments}

\subsection{Evaluation Framework}

We adopt the same indexing, retrieval, and evaluation framework as BEST-STD \cite{singh2025best} to
ensure a controlled comparison.
Each audio track in the speech archive is segmented into overlapping fixed-length
segments of duration $l$ with hop size $h$.
Each segment is tokenized into a discrete token sequence
$\mathbf{q}_{ij}=\{q_1,\ldots,q_T\}$, where $q_t \in \{1,\ldots,K\}$.

Since similarity between token sequences is computed using Jaccard similarity,
which does not explicitly encode temporal structure, we generate bigrams from the
token sequence to form
$\mathbf{B}_{ij}=\{(q_t,q_{t+1})\}_{t=1}^{T-1}$.
An inverted index $\mathcal{I}dx^{-1}$ is constructed by mapping each bigram to the set
of segments in which it appears:
\begin{equation}
\mathcal{I}dx^{-1}(b) = \{(i,j)\mid b \in \mathbf{B}_{ij}\}, \quad
b \in \{1,\ldots,K\}^2 .
\end{equation}

For a spoken query $\mathcal{Q}$, we tokenize it to obtain a bigram sequence
$\mathbf{Q}=\{b_1,\ldots,b_{\tilde{T}}\}$.
Retrieval is performed in two stages.
First, a coarse search identifies candidate segments that share at least one
bigram with the query:
\begin{equation}
\mathcal{C}oarse = \bigcup_{b \in \mathbf{Q}} \mathcal{I}dx^{-1}(b).
\end{equation}
Second, for each candidate segment $cand \in \mathcal{C}oarse$, we compute the maximum
Jaccard similarity between $\mathbf{Q}$ and all length-$\tilde{T}$ subsequences
of $cand$:
\begin{equation}
sim(cand) = \max_{t} Jaccard\big(\mathbf{Q}, cand[t:t+\tilde{T}-1]\big),
\end{equation}
where $Jaccard(\cdot,\cdot)$ denotes Jaccard similarity.
Candidates are ranked by $sim(cand)$ and the top-$k$ results are returned.

\begin{table}[th]
\caption{Discrete Token Consistency. Average Jaccard similarity between discrete token representations of paired spoken term utterances on LibriSpeech \texttt{train-clean-100} subset.}
\centering
\scriptsize
\begin{tabular}{lccc}
\toprule
\textbf{Model} & \textbf{Tokens} & \textbf{Unigram} & \textbf{Bigram} \\
\midrule
HuBERT-Base & 512  & 0.25 & 0.08 \\
HuBERT-Base & 1000 & 0.22 & 0.10 \\
WavLM-Base  & 512  & 0.33 & 0.16 \\
WavLM-Base  & 1000 & 0.28 & 0.14 \\
Encodec     & 1024 & 0.12 & 0.06 \\
Speech Tokenizer & 1024 & 0.44 & 0.24 \\
\midrule
 BEST-STD & 256 & 0.74 & 0.59 \\ 
BEST-STD & 512 & 0.73 & 0.57 \\ 
 BEST-STD & 1024 & 0.70 & 0.53 \\ 
  wav2tok & 256 & 0.80 & 0.72 \\ 
wav2tok  & 512 & 0.77 & 0.65 \\ 
wav2tok  & 1024 & 0.73 & 0.61 \\ 
 \midrule
\multicolumn{4}{c}{\textbf{Ours}} \\
 \midrule
 wav2tok 2.0 & 256 & 0.83 & 0.75 \\ 
wav2tok 2.0 & 512 & 0.81 & 0.71 \\ 
wav2tok 2.0 & 1024 & 0.79 & 0.68 \\ 
\bottomrule
\end{tabular}

\label{tab:Jaccard_Similarity}
\end{table}

\subsection{Datasets}

We train wav2tok 2.0 on the LibriSpeech~\cite{panayotov2015librispeech}
\texttt{train-clean-360} subset, using \texttt{test-clean} for model selection.
Retrieval is performed over the \texttt{train-clean-100} subset
($\sim$100 hours). Evaluation queries consist of two sets of 300 spoken terms
drawn from \texttt{train-clean-100} with no overlap with training utterances:
\emph{in-vocabulary} (IV) queries share lexical forms with training data but are
spoken by unseen speakers, while \emph{out-of-vocabulary} (OOV) queries contain
both unseen lexical forms and speakers.

To assess robustness under distributional shift, we additionally conduct
retrieval on the unseen TIMIT~\cite{garofolo1993timit} corpus, using
the train split (2,680 audio files) as the retrieval database and constructing IV
and OOV query sets of comparable size.

\subsection{Metrics}

We evaluate retrieval performance using standard QbE-STD metrics:
Mean Reciprocal Rank (MRR), Mean Average Precision (MAP), and Maximum Term
Weighted Value (MTWV).

\subsection{Baselines}

We compare wav2tok 2.0 against BEST-STD~\cite{singh2025best} and representative
general-purpose speech tokenizers, including HuBERT~\cite{hsu2021hubert},
WavLM~\cite{chen2022wavlm}, SpeechTokenizer~\cite{zhang2023speechtokenizer}, and
EnCodec~\cite{defossez2022high}. Token sequences for these models are obtained via
K-means clustering where applicable. We additionally evaluate conventional STD
systems based on frame-level features, including MFCCs~\cite{kamper2016deep},
bottleneck features~\cite{silnova2018but}, and posterior probabilities~\cite{abad2016exploiting},
combined with DTW-based matching. All baselines use the same indexing and retrieval
pipeline.

We also include wav2tok~\cite{banerjee2022wav2tok} as a baseline to isolate the
effect of the proposed DTW-aligned framewise token prediction loss. In this
setting, wav2tok shares the same training framework as wav2tok 2.0 but omits the
framewise prediction objective, relying solely on the CTC-based pairwise alignment
loss.

We do not evaluate LAST-STD~\cite{singh2025language}, which targets multilingual
settings and employs an optimal transport (OT) regularization~\cite{cuturi2013sinkhorn}
to address codebook utilization, nor BEST-STD~2.0~\cite{singh2025best2}, which
extends BEST-STD for noise robustness via OT-based codebook balancing. These
extensions are complementary to our alignment-based objectives and can be
straightforwardly incorporated into wav2tok 2.0-style training.

% We compare wav2tok 2.0 against BEST-STD and other representative speech tokenizers, including HuBERT, WavLM,
% SpeechTokenizer, and EnCodec, using token sequences extracted via K-means
% clustering where applicable.
% We also evaluate conventional STD systems based on frame-level
% features, including MFCCs, bottleneck features, and posterior probabilities,
% combined with DTW-based matching
% All baselines are evaluated using the same indexing and retrieval pipeline.

% \s

\begin{table*}[t]
\caption{Spoken Term Detection Performance. QbE-STD retrieval results on LibriSpeech \texttt{train-clean-100} and unseen TIMIT, reported using MAP, MRR, and MTWV for in-vocabulary and out-of-vocabulary queries.}

\centering
\scriptsize
\begin{tabular}{lcccc|ccc|ccc|ccc}
\toprule
\multirow{4}{*}{\textbf{Model} }& \multirow{4}{*}{\textbf{Tokens} }& \multicolumn{6}{c}{\textbf{LibriSpeech} \texttt{train-clean-100}} & \multicolumn{6}{c}{\textbf{TIMIT}} \\
\cmidrule(lr){3-8} \cmidrule(lr){9-14}
& & \multicolumn{3}{c}{\textbf{In-Vocabulary}} & \multicolumn{3}{c}{\textbf{Out-of-Vocabulary}} &\multicolumn{3}{c}{\textbf{In-Vocabulary}} & \multicolumn{3}{c}{\textbf{Out-of-Vocabulary}}  \\

\cmidrule(lr){3-5} \cmidrule(lr){6-8} \cmidrule(lr){9-11} \cmidrule(lr){12-14}
& & \textbf{MAP} & \textbf{MRR} & \textbf{MTWV} & \textbf{MAP} & \textbf{MRR} & \textbf{MTWV} & \textbf{MAP} & \textbf{MRR} & \textbf{MTWV} & \textbf{MAP} & \textbf{MRR} & \textbf{MTWV}  \\
\midrule
MFCC &  -  & 0.30 & 0.35 & 0.46 & 0.39 & 0.44 & 0.43 & 0.29 & 0.37 & 0.48 & 0.43 & 0.44 & 0.42 \\
Phone Posteriors &- & 0.42 & 0.44 & 0.50 & 0.47 & 0.50 & 0.48 & 0.41 & 0.44 & 0.52 & 0.41 & 0.43 & 0.47 \\
BNF & -  & 0.15 & 0.24 & 0.17 & 0.16 & 0.19 & 0.11 & 0.19 & 0.26 & 0.19 & 0.20 & 0.23 & 0.22 \\
HuBERT-Base & 512  & 0.22 & 0.25 & 0.40 & 0.22 & 0.23 & 0.51 & 0.20 & 0.20 & 0.32 & 0.25 & 0.26 & 0.31 \\
HuBERT-Base & 1000 & 0.18 & 0.20 & 0.33 & 0.22 & 0.21 & 0.31 & 0.19 & 0.17 & 0.23 & 0.22 & 0.22 & 0.16 \\
WavLM-Base  & 512  & 0.35 & 0.39 & 0.45 & 0.35 & 0.36 & 0.48 & 0.30 & 0.30 & 0.38 & 0.32 & 0.33 & 0.35 \\
WavLM-Base  & 1000 & 0.30 & 0.31 & 0.39 & 0.32 & 0.29 & 0.37 & 0.26 & 0.27 & 0.30 & 0.29 & 0.29 & 0.31 \\
Encodec     & 1024 & 0.15 & 0.16 & 0.23 & 0.16 & 0.16 & 0.26 & 0.08 & 0.08 & 0.13 & 0.03 & 0.02 & 0.16 \\
Speech Tokenizer & 1024 & 0.47 & 0.51 & 0.46 & 0.43 & 0.43 & 0.53 & 0.38 & 0.39 & 0.38 & 0.37 & 0.38 & 0.37 \\
\midrule
{BEST-STD} & 256  & 0.71 & 0.74 & 0.51 & 0.68 & 0.69 & 0.45 & 0.62 & 0.64 & 0.57 & 0.58 & 0.62 & 0.52 \\
{BEST-STD} & 512  & 0.73 & 0.78 & 0.56 & 0.70 & 0.71 & 0.51 & 0.61 & 0.66 & 0.63 & 0.59 & 0.64 & 0.55 \\
{BEST-STD} & 1024 & 0.65 & 0.70 & 0.61 & 0.64 & 0.65 & 0.54 & 0.57 & 0.62 & 0.62 & 0.55 & 0.59 & 0.58 \\

{wav2tok } & 256  & 0.82 & 0.85 & 0.59 & 0.78 & 0.79 & 0.52 & 0.71 & 0.74 & 0.66 & 0.67 & 0.71 & 0.60 \\
{wav2tok } & 512  & 0.80 & 0.86 & 0.61 & 0.77 & 0.78 & 0.56 & 0.67 & 0.72 & 0.69 & 0.65 & 0.70 & 0.60 \\
wav2tok & 1024 & 0.71 & 0.77 & 0.67 & 0.70 & 0.71 & 0.59 & 0.63 & 0.68 & 0.68 & 0.60 & 0.65 & 0.64 \\

 \midrule
\multicolumn{14}{c}{\textbf{Ours}} \\
 \midrule
{wav2tok 2.0} & 256  & 0.85 & 0.89 & 0.61 & 0.81 & 0.83 & 0.54 & 0.74 & 0.77 & 0.68 & 0.69 & 0.74 & 0.62 \\
{wav2tok 2.0 } & 512  & 0.86 & 0.90 & 0.66 & 0.82 & 0.84 & 0.60 & 0.72 & 0.78 & 0.74 & 0.69 & 0.75 & 0.65 \\
{wav2tok 2.0 } & 1024 & 0.78 & 0.84 & 0.74 & 0.77 & 0.78 & 0.65 & 0.69 & 0.75 & 0.75 & 0.66 & 0.71 & 0.70 \\

\bottomrule
\end{tabular}

% \vspace{-10pt}
\label{tab:STD}
\end{table*}

\subsection{Implementation Details}

Audio is segmented into fixed-length windows of $l=1$\,s with contextual padding,
a choice motivated by the observation that approximately $93\%$ of unique terms
in \texttt{train-clean-360} have durations shorter than 1\,s.
Each segment is converted into a 96-dimensional log-Mel spectrogram.

The encoder consists of four bidirectional Mamba \cite{gu2021efficiently} layers followed by a projection
to a 512-dimensional embedding space.
The total model contains approximately 4.7M trainable parameters.
We experiment with codebook sizes $K \in \{128,256,512,1024\}$.
The temperature parameter in the contrastive loss is fixed to $\tau=0.2$.
Stage I pretraining is conducted for 783 epochs using Adam with a learning rate of $5\times10^{-4}$. Stage II token alignment is conducted for a further 40 epochs. 

\section{Results}
\label{sec:results}
\subsection{Pairwise Token Consistency Analysis}

We evaluate pairwise token consistency using Jaccard similarity over unigram and
bigram token sets, where bigrams partially capture local order information and
are more sensitive to alignment-preserving tokenizations. As shown in
Table~\ref{tab:Jaccard_Similarity}, general-purpose tokenizers such as
HuBERT~\cite{hsu2021hubert}, WavLM~\cite{chen2022wavlm},
SpeechTokenizer~\cite{zhang2023speechtokenizer}, and
EnCodec~\cite{defossez2022high} exhibit relatively low consistency despite being
pretrained on large and diverse corpora,  reflecting that these models are trained as large-context encoders and have limited acoustic context in such short $\leq$ 0.5 s word crops. BEST-STD~\cite{singh2025best} improves
both unigram and bigram similarity via contrastive learning and vector
quantization with implicit DTW-based positive sampling. wav2tok 2.0 achieves the
highest consistency across all codebook sizes, with particularly strong gains in
bigram Jaccard similarity, indicating improved preservation of local sequential
structure. Ablation results in
Table~\ref{tab:Jaccard_Similarity} show that introducing explicit
CTC-based pairwise alignment (wav2tok~\cite{banerjee2022wav2tok}) yields
substantial improvements over BEST-STD \cite{singh2025best}, while the proposed DTW-aligned framewise
token prediction loss in wav2tok 2.0 further strengthens bigram-level consistency
through fine-grained cross-utterance agreement. While smaller codebooks yield
higher Jaccard scores at the cost of reduced discriminability, wav2tok 2.0
consistently outperforms both BEST-STD \cite{singh2025best} and wav2tok \cite{banerjee2022wav2tok} across all vocabulary sizes,
demonstrating that explicit pairwise alignment improves token stability. 
%without
%sacrificing scalability.

\subsection{Spoken Term Detection}

We evaluate downstream QbE-STD performance following the BEST-STD~\cite{singh2025best}
protocol, reporting MAP, MRR, and MTWV for in-vocabulary (IV) and
out-of-vocabulary (OOV) queries on LibriSpeech~\cite{panayotov2015librispeech}
\texttt{train-clean-100}, and additionally assess cross-dataset generalization on
the unseen TIMIT~\cite{garofolo1993timit} corpus (Table~\ref{tab:STD}). wav2tok 2.0
consistently outperforms all baselines across datasets, query types, and codebook
sizes, achieving the highest MAP and MRR for both IV and OOV queries and yielding
substantial gains over BEST-STD \cite{singh2025best}. Improvements are particularly pronounced for IV
queries, where explicit alignment promotes highly consistent tokenizations across
speakers, while gains on OOV queries indicate generalization beyond memorized
lexical items via transferable subword structure. General-purpose tokenizers
(HuBERT~\cite{hsu2021hubert}, WavLM~\cite{chen2022wavlm},
SpeechTokenizer~\cite{zhang2023speechtokenizer}, EnCodec~\cite{defossez2022high})
and conventional DTW-based baselines lag behind retrieval-oriented tokenizers,
reflecting the absence of alignment-driven training objectives, despite in some
cases being trained on much larger corpora. Ablation results show that introducing
explicit CTC-based pairwise alignment (wav2tok~\cite{banerjee2022wav2tok}) already
improves retrieval over BEST-STD \cite{singh2025best}, while the proposed DTW-aligned framewise token
prediction in wav2tok 2.0 further boosts MAP/MRR and consistently increases MTWV,
demonstrating the benefit of fine-grained alignment supervision. Although all
methods experience performance degradation under distribution shift on TIMIT \cite{garofolo1993timit},
wav2tok 2.0 remains strongest in both IV and OOV settings. Increasing
codebook size generally improves MTWV by enhancing discriminability, while mild
MAP/MRR degradation at large vocabularies reflects a robustness trade-off that
wav2tok 2.0 consistently mitigates relative to prior methods.

\section{Conclusion}
\label{sec:conclusion}

We propose \emph{wav2tok 2.0}, a scalable retrieval-oriented speech tokenizer that makes explicit pairwise alignment a first-class training signal. Built on BEST-STD \cite{singh2025best}, wav2tok 2.0 adds CTC-based sequence alignment with a novel DTW-aligned framewise prediciton objective, yielding more stable (especially bigram-consistent) tokenizations and consistent QbE-STD gains on LibriSpeech \cite{panayotov2015librispeech} and unseen TIMIT \cite{garofolo1993timit} over BEST-STD \cite{singh2025best}, wav2tok \cite{banerjee2022wav2tok}, and general-purpose tokenizers, without sacrificing efficiency. Future work includes integration of OT-based codebook balancing regularizers \cite{cuturi2013sinkhorn} and extension to multilingual, noisy, and long-form retrieval. Our speech tokenization frame-
work also shows promise for applications in developing speech LLMs \cite{chu2023qwen, hu2024wavllm}.
% \newpage

\section{Declaration of LLM Usage.} \label{llm_usage}

LLM is used only to aid or polish writing and does not impact the core methodology, scientific rigorousness, or originality of the research.

\bibliographystyle{IEEEtran}
\bibliography{mybib}

\end{document}